\documentclass[12pt,twoside]{article}
\usepackage{fleqn,espcrc1}

\def\lsim{\mathrel{\rlap{\lower4pt\hbox{\hskip1pt$\sim$}}
    \raise1pt\hbox{$<$}}}         %less than or approx. symbol
\def\gsim{\mathrel{\rlap{\lower4pt\hbox{\hskip1pt$\sim$}}
    \raise1pt\hbox{$>$}}}         %greater than or approx. symbol

\def\be{\begin{equation}}
\def\ee{\end{equation}}
\def\bq{\begin{eqnarray}}
\def\eq{\end{eqnarray}}

\title{Transverse polarization 
distributions
}

\author{Vincenzo Barone\address{Universit{\`a} di Torino and
        INFN, Sezione di Torino, 10125 Torino, Italy, \\
        and Universit{\`a} A. Avogadro, 15100 Alessandria, Italy }}

\begin{document}

\maketitle

\begin{abstract}
I present a brief update on the 
transverse polarization distributions, 
focusing on model calculations and phenomenological 
perspectives. 

\end{abstract}

\section{Introduction}

The transverse polarization (or transversity) distributions 
$\Delta_T q(x)$, introduced 20 years ago by Ralston and Soper 
\cite{RS} and studied in more detail in the last 
decade \cite{JJ}, are one of the three sets    
 of leading-twist quark and 
antiquark distribution 
functions -- the other two  are the momentum distributions
$q(x)$ and the helicity distributions $\Delta q(x)$. (Note that 
$\Delta_T q$ is often called $h_1$).

%All these quantities are contained in the 
% quark correlation matrix 
%\be
%\Phi_{ij} = \int {\rm d}^4 z \; {\rm e}^{i k \cdot z} \, 
%\langle P S \vert \bar \psi_j (0) \psi_i(z) \vert P S \rangle \,,  
%\label{corr}
%\ee
%where $P$ and $S$ are the momentum and the spin 
%of the proton, respectively,  and $k$ is the  momentum of the quark.  
%The Lorentz decomposition of (\ref{corr})
%reads ($\lambda$ is the proton helicity)
%\be
%\Phi \sim q(x) \, \slash P + \lambda \, \Delta q (x) \, \gamma_5 
%\slash P + \Delta_T q(x) \, \gamma_5 
%[\slash S_T, \slash P ]\;.   
%\label{corr2}
%\ee
Formally $\Delta_T q$ is given by 
%From this one can extract the 
% transverse polarization distributions ($n$ is a 
%light-like four-vector) 
\be
\Delta_T q(x) =
\frac{1}{\sqrt{2} P^+}\int \frac{{\rm d}\alpha}{2 \pi}\,
{\rm e}^{i \alpha \, x}
\langle P S|\psi_+ ^\dagger(0)
\gamma_\perp\gamma_5 \psi_+ (\alpha n)| P S\rangle\;, 
\label{defh1}
\ee
where 
 $P$ and $S$ are the momentum and the spin 
of the proton, respectively, $n$ is a null vector such that 
$n \cdot P =1$, and $\psi_+={1\over 2}\gamma^-\gamma^+\psi$. 
 The antiquark distributions are obtained from (\ref{defh1}) 
by exchanging 
 $\psi$ with $\psi^\dagger$. 
%Note that in eq.~(\ref{defh1})
% only the `good'
%light-cone components of the fields, 
%$\psi_+={1\over 2}\gamma^-\gamma^+\psi$, 
%appear, as it is always the case for 
% leading twist quantities. 

Inserting a complete
set of intermediate 
states $\{ \vert X \rangle \}$ and using the 
Pauli--Lubanski projectors 
${\cal P}_\perp^{\uparrow \downarrow}
=\frac{1}{2} (1 \pm \gamma_{\perp} \gamma_5)$  
one gets 
\be
\Delta_T q(x) = \frac{1}{\sqrt{2}} \, 
\sum_X \{ \vert\langle P S |{\cal P}_\perp^{\uparrow} 
\psi_+(0)\vert X \rangle\vert^2-
\vert\langle PS 
|{\cal P}_\perp^{\downarrow}\psi_+(0)\vert X \rangle\vert^2 
\} \, \delta[(1-x)P^+ - p_X^+] \,.
\label{2}
\ee
which clearly shows the probabilistic meaning of $\Delta_T q$ 
in the transverse polarization basis: $\Delta_T q(x)$ is the number 
density of quarks with  momentum fraction $x$ and transverse 
polarization $\uparrow$ minus the number density of quarks 
with the same momentum and 
transverse polarization $\downarrow$, in a transversely 
polarized hadron. 
 In the helicity basis $\Delta_T q$ is non diagonal 
and hence has no probabilistic interpretation.  

Being a chirally odd distribution, 
 $\Delta_T q$  is not 
measurable in deep inelastic scattering.  This makes it 
quite an elusive quantity. 
At present
we have no experimental information on it.  
That is why    model calculations and other 
nonperturbative studies are particularly useful.

%%%%%%%%%%%%%%%%%%%%%%%%%%%%%%%%%%%%%%%%%%%%%%%%%%
\section{Models}

 The transverse polarization distributions
have been calculated in a large number of models:   
{\it i)} bag model \cite{jaffe,stratmann,scopetta}, 
{\it ii)} chromodielectric model \cite{noi1}, 
{\it iii)} chiral quark soliton and NJL model \cite{pobylitsa,gamberg}, 
{\it iv)} light-cone models \cite{ma,traini}, {\it v)} 
spectator model \cite{mulders,suzuki}.

Many of these calculations show  that,  
at  small $Q^2$ ($\lsim 0.5$ GeV$^2$),
 $\Delta_T q$  is not very different from   
$\Delta q$, 
at least for $x \gsim 0.1$. At low $x$ the situation 
is more controversial: some 
 models \cite{pobylitsa,traini} predict a sensible difference 
between the two  distributions. A definite conclusion 
cannot be drawn 
since the various models are valid at different 
scales and it is 
 known that the QCD evolution 
induces a difference between $\Delta_T q$ and $\Delta q$
which is relevant especially at low $x$ \cite{barone}.

As for the tensor 
 charges
\be
\delta q = \int_0^1 {\rm d} x \, \, [\Delta_T q (x) 
- \Delta_T \bar q(x) ]\;, 
\label{deltaq}
\ee
in addition to the predictions of  
 the models listed above, there are  other  
nonperturbative estimates: by QCD sum rule methods \cite{he},  
and by lattice QCD \cite{aoki}. 

A rough (and personal)  average of all model  results is 
\[
\delta u \sim 1.0 \pm 0.2\;, \;\;\;\;
\delta d \sim -0.3 \pm 0.1 \;, \;\;\;
{\rm at} \; \; Q^2 \sim 2 \, {\rm GeV}^2
\]
where the error does not account for the intrinsic 
uncertainty of each model, but represents only the 
range spanned by the various results. 
For comparison, the lattice finding \cite{aoki} is 
\[
\delta u = 0.84 \;, \;\;\;\;
\delta d = -0.23 \;, \;\;\;
{\rm at} \; \; Q^2 \sim 2 \, {\rm GeV}^2\;. 
\]

Note that $\delta u$ and $\delta d$ are, in absolute value, 
only slightly smaller than the nonrelativistic expectations
($ \delta u_{NR} = 4/3$, $\delta d_{NR} = -1/3$).

%%%%%%%%%%%%%%%%%%%%%%%%%%%%%%%%%%%%%%%%%%%%%%%%%%

\section{Possible measurements}

As already  mentioned, the transverse polarization 
distributions cannot be measured in inclusive DIS. 
To extract $\Delta_T q$ one needs either two hadrons 
 in the initial state (hadron-hadron collisions), 
or one hadron in the initial state and one in the final state 
(semiinclusive deep inelastic scattering).   

The measurement of $\Delta_T q$ in proton--proton collisions
is part of the physics program 
of the experiments
at RHIC \cite{RHIC} and of the proposed HERA-$\vec N$ 
project \cite{HERAN}.

Among the possible $pp$ initiated 
processes one can make a 
selection choosing those which are expected to yield the 
largest spin asymmetry. 
Since there is no  gluon transversity distribution \cite{JJ}, 
all processes dominated at the partonic level  by 
 $qg$ or $gg$ scattering 
 produce a very small transverse asymmetry 
\cite{ji}. Hence 
the most promising reaction is  Drell-Yan lepton pair 
production with two transversely polarized beams. 
The relevant observable is 
 the double--spin transverse
asymmetry
\be
A_{TT} = 
\frac{d \sigma_{\uparrow \uparrow} - 
d \sigma_{\uparrow \downarrow}}{d \sigma_{\uparrow \uparrow} +
d \sigma_{\uparrow \downarrow}}\,,  
\label{att}
\ee
which depends on  the product ($A$ and $B$ are the two protons)
\be
\Delta_T q (x_A) \, \Delta_T q(x_B)\;. 
\label{dy}
\ee

The Drell-Yan $A_{TT}$ has been calculated at leading order 
\cite{noi1,noi2} and  next-to-leading order \cite{vogelsang}.  
In \cite{noi1,noi2}  
 $\Delta_T q = \Delta q$ was assumed at a very low scale
(the input $\mu^2$ of the  GRV distributions). The authors   
of \cite{vogelsang}, instead,  set $\vert \Delta_T q \vert 
= 2 \, (q + \Delta q)$ at the GRV scale, assuming 
the saturation of Soffer's inequality. This yields the maximal 
value for $A_{TT}$. 
Summarizing the results of these calculations 
we can say that at RHIC energies 
($\sqrt{s} > 100$ GeV) one expects for the double-spin 
asymmetry, integrated over the invariant mass $M^2$ of the 
dileptons
\be
A_{TT} \sim (1-2) \% \;, \;\;\; {\rm at} \;\; {\rm most}\;.
\label{att2}
\ee
It is quite interesting to note that,  
as $\sqrt{s}$ gets lower, the asymmetry tends to increase, 
as it was first pointed out in \cite{noi1,noi2}. 
Thus at the HERA-$\vec N$ energies ($\sqrt{s} = 40$ GeV) 
$A_{TT}$ can reach a value of $\sim (3-4) \%$,  
which 
 should be measurable within the expected statistical 
errors for that experiment \cite{vogelsang}.

Let us turn now to semiinclusive DIS on a transversely 
polarized proton. There are three candidate reactions 
for determining $\Delta_T q$ at leading twist.  

Detecting a transversely polarized hadron $\vec h$ 
(e.g., a $\Lambda$) in the 
final state,
\be
e \, \vec p \rightarrow e \, \vec h \, X\; , 
\ee
one measures the product 
\be
\Delta_T q(x) \, H_1^q(z) \;, 
\ee
where $H_1^q$ is a chirally odd leading-twist 
fragmentation function. 
In principle there is no reason why $H_1$ should be 
much smaller than the unpolarized fragmentation function
$D_1$. The model calculation of \cite{mulders}
gives for instance 
$H_1^u /D_1^u \sim 0.5$ and  $H_1^d /D_1^d \sim -0.2$. 

The second relevant reaction is semiinclusive DIS with 
an unpolarized final hadron
\be
e \, \vec p \rightarrow e \,  h \, X\;.  
\ee
In this case $\Delta_T q$ might appear as a consequence of 
the Collins effect \cite{collins} (a T-odd 
contribution arising from final state interactions). 
Here one measures 
\be
\Delta_T q(x) \, H_1^{\perp q}(z) \;, 
\ee
where $H_1^{\perp q}$ is a T-odd leading-twist fragmentation 
function. The estimate of  $H_1^{\perp q}$
presented in \cite{anselmino} and based on the analysis 
of $p p$ reactions  shows that 
this quantity is non negligible only at  high $z$. 

A third way to extract $\Delta_T q$ from semiinclusive 
DIS has been explored in \cite{tang}. The idea 
is to study the process
\be
e \, \vec p \rightarrow e \, h_1 \, h_2 \,   X 
%&\hookrightarrow& h_1 \, h_2 \, X 
\ee 
where $h_1, h_2$ are two mesons in a correlated state which  
is the superposition of two resonances $h,h'$
\be
\vert h_1 \, h_2 \rangle = {\rm e}^{i \delta} \, 
\vert h \rangle + {\rm e}^{i \delta'} \, \vert h' \rangle\;. 
\ee
For instance,  $h_1,h_2 = \pi^+, \pi^-$ 
and $h,h' = \sigma,\rho$. In this reaction one 
measures 
\be
\sin{\delta} \, \sin{\delta'} \, \sin{(\delta-\delta')} 
\, \Delta_T q(x) \, I_q(z) 
\label{pipi}
\ee
where $I_q(z)$ is the $hh'$ interference fragmentation 
function. Nothing is known at present about this quantity.  

From this  sketchy presentation of the phenomenological 
perspectives it should be clear that, whereas
the Drell-Yan process allows to determine $\Delta_T q$  
in a  clean way, semiinclusive DIS 
is characterized by the presence of fragmentation functions which 
are little known and, in some cases, 
are expected to be 
rather small.

\vspace{0.5cm}
It is a pleasure to  thank A.~Drago for a long   
collaboration work on this subject and M.~Anselmino 
for useful discussions.

\end{document}